\pdfoutput=1

\documentclass[prl,twocolumn,twoside,preprintnumbers,superscriptaddress,nofootinbib]{revtex4-1}

\usepackage{amsmath}
\usepackage{amssymb}
\usepackage{mathrsfs}
\usepackage{graphicx}
\usepackage[hyperfootnotes=false]{hyperref}
\usepackage{xspace}
\usepackage[dvipsnames]{xcolor}
\usepackage[utf8]{inputenc}
\usepackage[normalem]{ulem}

\newcommand{\eq}[1]{Eq.~\eqref{eq:#1}}

\newcommand{\fig}[1]{Fig.~\ref{fig:#1}}

\newcommand{\ord}[1]{{\mathcal O}(#1)}
\newcommand{\vecb}[1]{\mbox{\boldmath $#1$}}
\newcommand{\sandwich}[3]{\left< #1 \right | #2 \left | #3 \right >}

\newcommand{\nn}{\nonumber}

\newcommand{\df}{\mathrm{d}}
\newcommand{\img}{\mathrm{i}}
\newcommand{\sdt}{\!\cdot\!}

\newcommand{\al}{\alpha}
\newcommand{\bt}{\beta}
\newcommand{\ga}{\gamma}

\newcommand{\de}{\delta}

\newcommand{\si}{\sigma}

\newcommand{\thh}{\theta}

\newcommand{\ze}{\zeta}

\newcommand{\cL}{{\mathcal L}}

\newcommand{\cC}{{\mathcal C}}

\newcommand{\bnslash}{\bar{n}\!\!\!\slash}
\newcommand{\bn}{{\bar{n}}}

\newcommand{\MSbar}{\ensuremath{\overline{\text{MS}}}}

\newcommand{\WTA}{\mathrm{WTA}}

\newcommand{\axis}{\mathrm{axis}}

\newcommand{\LzeroQTR}{\cL^0_+\Big(q,\frac{QR}{2}\Big)}
\newcommand{\LoneQTR}{\cL^1_+\Big(q,\frac{QR}{2}\Big)}

\newcommand{\LB}{\ln\Big(\frac{|\vecb{b}|^2\mu^2}{4e^{-2\ga_E}}\Big)}

\allowdisplaybreaks[2]

\begin{document}


\preprint{\vbox{\hbox{Nikhef 2018-034}}}

\title{Transverse momentum dependent distributions with jets}

\author{Daniel Gutierrez-Reyes}

\affiliation{Departamento de F\'isica Te\'orica, Universidad Complutense de Madrid (UCM), E-28040 Madrid, Spain}

\author{Ignazio Scimemi}

\affiliation{Departamento de F\'isica Te\'orica, Universidad Complutense de Madrid (UCM), E-28040 Madrid, Spain}

\author{Wouter J.~Waalewijn}

\affiliation{Institute for Theoretical Physics Amsterdam and Delta Institute for Theoretical Physics, University of Amsterdam, Science Park 904, 1098 XH Amsterdam, The Netherlands}
\affiliation{Nikhef, Theory Group, Science Park 105, 1098 XG, Amsterdam, The Netherlands}

\author{Lorenzo Zoppi}

\affiliation{Institute for Theoretical Physics Amsterdam and Delta Institute for Theoretical Physics, University of Amsterdam, Science Park 904, 1098 XH Amsterdam, The Netherlands}
\affiliation{Nikhef, Theory Group, Science Park 105, 1098 XG, Amsterdam, The Netherlands}

\begin{abstract}
We investigate the use of jets to measure transverse momentum dependent distributions (TMDs). The example we use to present our framework is the dijet momentum decorrelation at lepton colliders. Translating this momentum decorrelation into an angle $\theta \ll 1$, we analyze the factorization of the cross section for the cases $\theta \gg R$, $\theta \sim R$ and $\theta \ll R$, where $R$ is the jet radius. Critically, for the Winner-Take-All axis, the jet TMD has the same double-scale renormalization group evolution as TMD fragmentation functions for all radii $R$. TMD fragmentation functions in factorization theorems may then simply be replaced by the jet TMDs we calculate, and all ingredients to perform the resummation to next-to-next-to-leading logarithmic accuracy are available. Our approach also applies to semi-inclusive deep inelastic scattering (SIDIS), where a jet instead of a hadron is measured in the final state, and we find a clean method to probe the intrinsic transverse momentum of quarks and gluons in the proton that is less sensitive to final-state nonperturbative effects. 
\end{abstract}

\maketitle

{\it Introduction --}
The precision of the large hadron collider (LHC) and the advent of the electron-ion collider (EIC) have focused the attention on differential cross sections that probe transverse momentum dependent distributions (TMDs). These  non-perturbative momentum distributions of quarks and gluons in hadrons (TMD parton distribution functions), and conversely of hadrons fragmenting from quarks or gluons (TMD fragmentation functions), are expressed in terms of a collinear momentum fraction and a momentum transverse to it. TMDs are process independent, entering in factorization theorems for processes like Drell-Yan, semi-inclusive deep inelastic scattering (SIDIS) and the production of two hadrons in $e^+e^-$ collisions~\cite{Collins:1981uk,Collins:1984kg,Meng:1995yn,Becher:2010tm,Collins:2011zzd,GarciaEchevarria:2011rb,Chiu:2012ir,Echevarria:2014rua}. 

We will concentrate on transverse momentum dependent distributions involving jets. This is a natural extension for the LHC, where measurements generically involve jets. The measurement of dijets is also part of the program at the relativistic heavy ion collider (RHIC). At the EIC jets are possibly measured, so that  the question of  how one can extract TMDs using  jets  in a SIDIS experiment is relevant, see e.g.~Refs.~\cite{Boer:2016fqd,Page:2017pcx}.

One possible direction is to study hadrons inside jets. The factorization analysis in Refs.~\cite{Kang:2017glf,Kang:2017yde} showed that the standard TMD fragmentation functions enter when the transverse momentum of a hadron is measured with respect to the standard jet axis (SJA). If instead the Winner-Take-All (WTA) axis~\cite{Salam:WTAUnpublished,Bertolini:2013iqa} is used, a new and rather different TMD is obtained: it is insensitive to soft radiation and has a DGLAP-like evolution equation~\cite{Neill:2016vbi}.

Here we explore a different direction, by considering the transverse momentum of the jet itself. The example that will be discussed in detail is the dijet momentum decorrelation, defined in \eq{obs}. This is closely related to the azimuthal decorrelation measured at the Tevatron~\cite{Abazov:2004hm}, RHIC~\cite{Adams:2006yt} and LHC~\cite{Khachatryan:2016hkr,Aaboud:2018hie}, and calculated to next-to-leading logarithmic accuracy in Refs.~\cite{Banfi:2008qs,Sun:2014gfa,Chen:2016jfu}. These calculations use the SJA and treat $\theta \ll R$, and Sudakov logarithms they resum differ from the standard TMD case that we are interested in. 

We are particularly interested in how TMD jet measurements can shed light on the intrinsic transverse momentum of quarks and gluons in the proton. To this end we extend our framework to SIDIS, with a hadron in the initial state, but where the final state is a jet. We will present the relevant factorization theorem in \eq{SIDIS}. The use of jets provides a clean way to extract the intrinsic transverse momentum distributions inside the  initial hadron. 
Contrary to the case of fragmentation functions, the momentum fraction of a jet is perturbatively calculable. The nonperturbative effect on the transverse momentum is formally of the same size, but we find it to be more suppressed for jets with the WTA axis. All ingredients necessary for resummation at next-to-next-to-leading logarithmic accuracy in our framework are now available. 

We will consider the anti-$k_t$ jet algorithm~\cite{Cacciari:2008gp}, exploring the dependence on the jet energy, radius $R$ and choice of jet axis (SJA or WTA). We are particularly interested in the conditions for which the jet TMDs have the same evolution as TMD fragmentation functions, as this puts them on equal footing. The evolution of TMDs can be derived from the soft function
\begin{align} \label{eq:softf}
S(\vecb b)=
\frac{1}{N_c}{\rm Tr}_c
\sandwich{0}{\!\bigl[S_n^{T\dagger} \tilde S_\bn^T \bigr](0^+,0^-,\vecb b_\perp)
\bigl[\tilde S^{T\dagger}_\bn S_n^T\bigr](0)\!}{0},
\end{align}
where $S_n^T$ and ${\tilde S}_\bn^T$ denote soft Wilson lines along (almost) back-to-back light-cone directions $n$ and $\bn$, including a transverse Wilson line~\cite{Bassetto:1984dq,Hautmann:2007uw,Cherednikov:2007tw,Cherednikov:2009wk,Idilbi:2010im,GarciaEchevarria:2011md}. TMD calculations involve rapidity divergences, that require a regulator in addition to dimensional regularization. The soft function in \eq{softf} has been calculated to higher orders in perturbative QCD in several rapidity regularization schemes~\cite{Echevarria:2015byo,Luebbert:2016itl,Li:2016axz,Li:2016ctv,Vladimirov:2016dll}, and it is responsible for the double scale renormalization group evolution typical of 
TMDs~\cite{Catani:2000vq,Becher:2010tm,Collins:2011zzd,GarciaEchevarria:2011rb,Aybat:2011zv,Chiu:2012ir,pippo,Catani:2013tia,Scimemi:2018xaf}. 
As it is clear, \eq{softf} is independent of any jet specifics, like $R$ or the jet algorithm, and in the following we will see under which conditions this soft function can also be used when dealing with jets, instead of hadrons.

We will first discuss the factorization theorems for the momentum decorrelation, then treat the renormalization and resummation, calculate the TMD jet function, and conclude with a discussion of the implications and interesting extensions of our work.

\begin{figure*}[tb]
\centering
\includegraphics[width=0.8\textwidth]{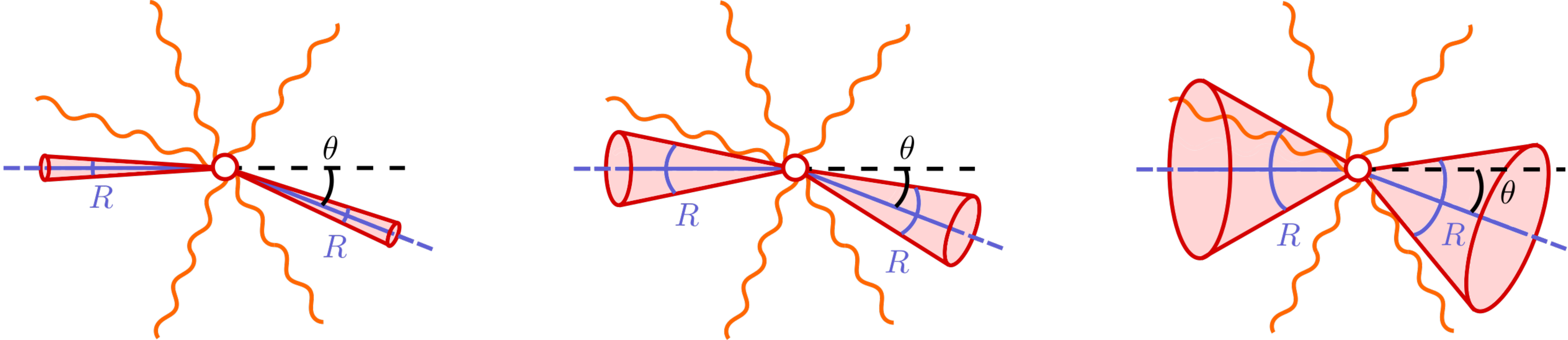}%
\caption{Various hierarchies between the jet radius $R$ and angular decorrelation $\theta$, $\theta \gg R$ (left)$,\theta \sim R$ (middle), and $\theta \ll R$ (right). For $\theta \gg R$ the dependence on the jet axis vanishes, while for $\theta \ll R$ it even modifies the factorization structure.}
\label{fig:TMDjetRegimes}
\end{figure*}

{\it Factorization --} We consider the momentum decorrelation of inclusive dijet production in $e^+e^-$ collisions. Inclusive means that each pair of jets contributes, with soft jets kept at bay by measuring the energy fraction of the jets $z_i = 2E_i/Q$, where $Q$ is the center-of-mass energy. The momentum decorrelation 
is\footnote{The axis with respect to which the transverse momenta $\vecb{p}_i$ are measured can be any axis close to the jets, as different choices only affect the $\vecb{q}^2/Q^2$ power corrections.}
\begin{align} \label{eq:obs}
  \vecb{q} = \frac{\vecb{p}_1}{z_1}  + \frac{\vecb{p}_2}{z_2}\ .
\end{align}
In direct analogy to the hadron case, we divide the transverse momentum $\vecb{p}_i$ of each jet by its energy fraction. Then $\vecb{q}$ has a direct correspondence to the dijet angular decorrelation $\theta \approx \tan \theta = 2|\vecb{q}|/Q$, where in the absence of additional QCD radiation $\theta=0$.
The parameter $\theta$ is generally assumed to be small, however it can compete with other small quantities like the jet radius $R$, leading us to study each case in Fig.~\ref{fig:TMDjetRegimes} separately.

First we consider the case $\theta \sim R \ll 1$, for which the interplay between $\theta$ and $R$ must be taken into account in the jet TMD. However, since $R \ll 1$, the wide-angle soft radiation does not resolve individual collinear emissions in the jet, so the soft function is the same as for TMD fragmentation. This leads to, 
\begin{align}
\label{eq:CS}
   \frac{\df \si_{(ee\rightarrow J\! J\! X)}}{\df z_1\, \df z_2\, \df \vecb{q}} &= H(Q^2,\mu)\!\!
   \int \!\! \frac{\df \vecb{b}}{(2\pi)^2} e^{-\img \textit{\textbf{b}} \cdot \textit{\textbf{q}}}
   J_q^\axis(z_1, \vecb{b}, QR,\mu, \zeta_1)
   \nn \\ & \quad \!\times\!
   J_{\bar q}^\axis(z_2, \vecb{b}, QR,\mu, \zeta_2)
   \biggl[1 \!+\! \mathcal{O}\Bigl(\frac{\vecb{q}^2}{Q^2}\Bigr)\biggr].
\end{align}
The hard function $H$ encodes the short-distance scattering, $e^+e^- \to q \bar q$, and is the same as for TMD fragmentation. 
The quark TMD jet function $J_q^{\rm axis}$ encodes the inclusive production of jets in one direction, where axis is either WTA or SJA, and $\vecb{b}$ is Fourier conjugate to the jet transverse momentum. In Soft-Collinear Effective Theory \cite{Bauer:2000ew,Bauer:2000yr,Bauer:2001ct,Bauer:2001yt} it is defined as \footnote{Note that our TMD jet function $J_q^{\rm axis}$ differs from a TMD fragmentation function by an overall factor of $z^2$.}
\begin{align} \label{eq:Jaxis_def}
J_q^{\rm axis} &=\frac{z}{2N_c}\text{Tr}\Big[\frac{\bnslash}{2} 
\langle 0|\big[\delta \big(\bn \sdt p_J/z-\bar n \cdot P\big) 
e^{-\img \textit{\textbf{b}}\cdot\textit{\textbf{P}}_\perp} \chi_n(0) \big]
\nn \\ & 
\quad
\times
\sum_X |J_{\text{anti-}k_t,R}^{\rm axis} X\rangle \langle J_{\text{anti-}k_t,R}^{\rm axis}X|\bar\chi_n(0)|0\rangle\Big].
\end{align}
Here $\chi_n$ is the collinear quark field in the light-like direction $n^\mu$, $\bn^\mu$ is a conjugate light-like vector with $\bar n\cdot n = 2$, $P$ is the momentum operator, and the delta function and exponential encode the measurement of the large- and transverse-momentum components.\footnote{Note that the \emph{total} jet momentum is not integrated over, and has zero transverse momentum for our choice of frame.} 
We have eliminated the soft function in \eq{softf} from \eq{CS}, by absorbing a square-root of it into each jet function.
 The jet depends on the jet algorithm (anti-$k_t$), radius $R$ and choice of jet axis, but the anomalous
  dimension  of the jet function does not. The logarithms of $\vecb{q}^2/Q^2 \sim \theta^2 \sim R^2$ are resummed by evaluating each ingredient at its natural $\mu$ and $\zeta$ scale and evolving them to a final scale using the $\mu$ and $\zeta$ renormalization group evolution, as will be discussed in more detail below.

Secondly, for $R \ll \theta \ll 1$ additional large logarithms of $\vecb{q}^2/(QR)^2 \sim (\theta/R)^2$ are induced. The form of the cross section is the same as in Eq.~(\ref{eq:CS}) with the substitution
\begin{align} \label{eq:smallTheta}
  J_i^\axis(z, \vecb{b}, QR, \mu, \zeta)&=
  \sum_j\! \int\! \frac{\df z'}{z'}\, \big[(z')^2\mathbb{C}_{i \to j}(z', \vecb{b}, \mu, \zeta)\big]
  \nn \\ & \quad \times
   {\cal J}_j\Big(\frac{z}{z'}, QR,\mu\Big)\,\big[1 \!+\! \mathcal{O}(\vecb{b}^2 Q^2R^2)\big]\,.
 \end{align}
The axes choice no longer affects the jet function because of the limit $R \ll \theta$. The semi-inclusive jet functions ${\cal J}_j$ describe inclusive production of jets, and are given to $\ord{\al_s}$ in Refs.~\cite{Kang:2016mcy,Dai:2016hzf}.\footnote{To use the expressions in Ref.~\cite{Kang:2016mcy}, note that $\omega_J = 2z' Q$.}  The matching coefficients $\mathbb{C}_{i\to j}$ are the same as for TMD fragmentation, and are given to $\ord{\al_s^2}$ in Refs.~\cite{Echevarria:2015usa,Echevarria:2016scs}. The fact that the same coefficients appear is not surprising, since in the fragmentation limit $ R\to 0$, the semi-inclusive jet function reduces to the fragmentation function summed over hadron species $h$, $ {\cal J}_j(z, QR,\mu) \to \sum_h d_{j \to h}(z,\mu)$~\cite{Kang:2017mda}.

Last of all we consider $\theta \ll R$, where the choice of jet axis directly affects the factorization theorem. This case is probably the most relevant for studying the small intrinsic transverse momentum of quarks and gluons in the proton using SIDIS. Interestingly, for the Winner-Take-All (WTA) axis we find,
\begin{align} \label{eq:wta}
  J_i^{\rm WTA}(z, \vecb{b}, QR, \mu, \zeta)
  &= \de(1-z)\, {\mathscr{J}}_i^{\rm WTA} (\vecb{b}, \mu, \zeta)
  \nn \\ & \quad \times
  \Big[1 + \mathcal{O}\Big(\frac{1}{\vecb{b}^2 Q^2R^2}\Big)\Big]
\,,\end{align}
where 
\begin{align} \label{eq:scriptJ}
{\mathscr{J}}_q^{\rm WTA}(\vecb{b},\mu,\ze)&=\frac{1}{2N_c(\bn \sdt p_J)}\text{Tr}\Big[\frac{\bnslash}{2} 
\langle 0|\big[
e^{-\img \textit{\textbf{b}}\cdot\textit{\textbf{P}}_\perp} \chi_n(0) \big]
\nn \\ & 
\quad
\times
|J_{\text{anti-}k_t}^\WTA \rangle \langle J_{\text{anti-}k_t}^\WTA|\bar\chi_n(0)|0\rangle\Big]
\,.\end{align}
The jet function ${\mathscr{J}}_q^{\rm WTA}$ differs from $J_q^{\rm WTA}$ in \eq{Jaxis_def}, because all collinear radiation has been clustered into the jet, implying that there is no dependence on the jet radius in this limit.

 Although the factorization in \eq{CS} was derived for $R\ll1$, we find that for the WTA axis it also holds for $\theta \ll R\sim 1$, as we now discuss: for the collinear radiation with typical angle $\theta$ the jet boundary seems infinitely far away, so it does not depend on $R$, as in \eq{scriptJ}. The soft radiation \emph{does} resolve the jet boundary, but since it does not affect the position of the WTA axis, there is no distinction between soft radiation inside or outside of the jet\footnote{We assume $z_i$ is measured in sufficiently large bins, as $1\!-\!z_i \ll 1$ \emph{is} sensitive to whether soft radiation is in or outside the jet.}. We thus obtain the same soft function as before, which accounts for the total recoil due to soft radiation.

By contrast, in the same $\theta \ll R$ limit, this statement is \emph{not} true for the SJA. The SJA is aligned with the total momentum in a jet, implying that the total transverse momentum $\vecb{q}$ is only sensitive to soft radiation outside the jets. Assuming $R \sim 1$, hard splittings with typical angle $R$ are allowed inside the jet, each generating a Wilson line sourcing this soft radiation. Thus we can describe this cross section by~\cite{Becher:2015hka} (see also Refs.~\cite{Larkoski:2015zka,Caron-Huot:2015bja})
\begin{align}
  \frac{\df \si^{\rm SJA}_{(ee\rightarrow J\! J\! X)}}{ \df \vecb{q}} 
 &= 
    \sum_{m=2}^\infty {\rm Tr}_c[ \mathcal{H}_m(Q, \{n_i\},R,\mu) 
    \nn \\ & \quad 
    \otimes \mathcal{S}_m(\vecb{q},\{n_i\},R,\mu)] \bigg[1 + \mathcal{O}\Big(\frac{\vecb{q}^2}{Q^2}\Big)\bigg]
,\end{align}
where the trace is over color indices and $\otimes$ denotes integrals over the directions $n_i$ (with $i=1, \dots, m$) of hard splittings inside the jets. Instead of the rather simple result in \eq{wta}, the observable is now intrinsically sensitive to non-global logarithms~\cite{Dasgupta:2001sh}.

{\it Renormalization group evolution --}  The double scale RG equations for the jet TMDs are the same as for the TMD fragmentation functions
\begin{align}
\mu^2 \frac{\df}{\df\mu^2} J^{\rm axis}_{i}(z,\vecb{b}, QR ,\mu,\zeta)&=\frac{1}{2}\gamma^i_F(\mu,\zeta)
J^{\rm axis}_{i}(z,\vecb b, QR,\mu,\zeta),\nn
\\\label{th:evol_zeta}
\zeta \frac{\df}{\df\zeta} J^{\rm axis}_{i}(z,\vecb b, QR,\mu,\zeta)&=-\mathcal{D}^i(\mu,\vecb b)J^{\rm axis}_{i}(z,\vecb b, QR,\mu,\zeta),
\end{align}
and are independent of axis. For the $\theta \ll R$ limit, $\mathscr{J}^\WTA$ has the same evolution, but there is no analogue for the SJA.
The TMD anomalous dimensions $\gamma_F$ and $\mathcal{D}$ are known up to $\ord{\al_s^3}$~\cite{Moch:2005id,Li:2016ctv,Vladimirov:2016dll}.
Solving these RG equations is straightforward in impact parameter space,
\begin{align}\label{th:TMD_evol}
&J^{\rm axis}_{i}(z,\vecb b, QR,\mu_f,\zeta_f)
\nn \\
&\quad =\exp\biggl[\int_{(\mu_i,\zeta_i)}^{(\mu_f,\zeta_f)} \biggl(\gamma^i_F(\mu,\zeta)\frac{\df\mu}{\mu} -\mathcal{D}^i(\mu,\vecb b)\frac{\df\zeta}{\zeta}\biggr)\biggr]
\nn\\ & \qquad \times 
J^{\rm axis}_{i}(z,\vecb b, QR,\mu_i,\zeta_i).
\end{align}
A study of evolved jet TMDs and their phenomenology will be presented in Ref.~\cite{nos}. 

{\it TMD jet function --}
The final point we discuss, is the NLO calculation of the TMD jet function. It also receives nonperturbative corrections, but these are suppressed by powers of $\vecb b^2 \Lambda_{\rm QCD}^2$. We use dimensional regularization (in the $\MSbar$ scheme) and the modified $\de$ regulator, which alters the Wilson lines in the operator definition~\cite{Chiu:2009yx,Echevarria:2015usa}. With the $\de$ regulator, the overlap of soft and collinear modes~\cite{Manohar:2006nz} coincides with the soft function in \eq{softf}. We therefore account for both the absorption of the soft function and this overlap by including a factor $S^{-\frac{1}{2}}$ in the TMD jet function. This operation replaces rapidity divergences by logarithms of the rapidity scale $\ze$~\cite{Echevarria:2016scs}.

\begin{figure*}[tb]
{\centering
\includegraphics[width=0.9\textwidth]{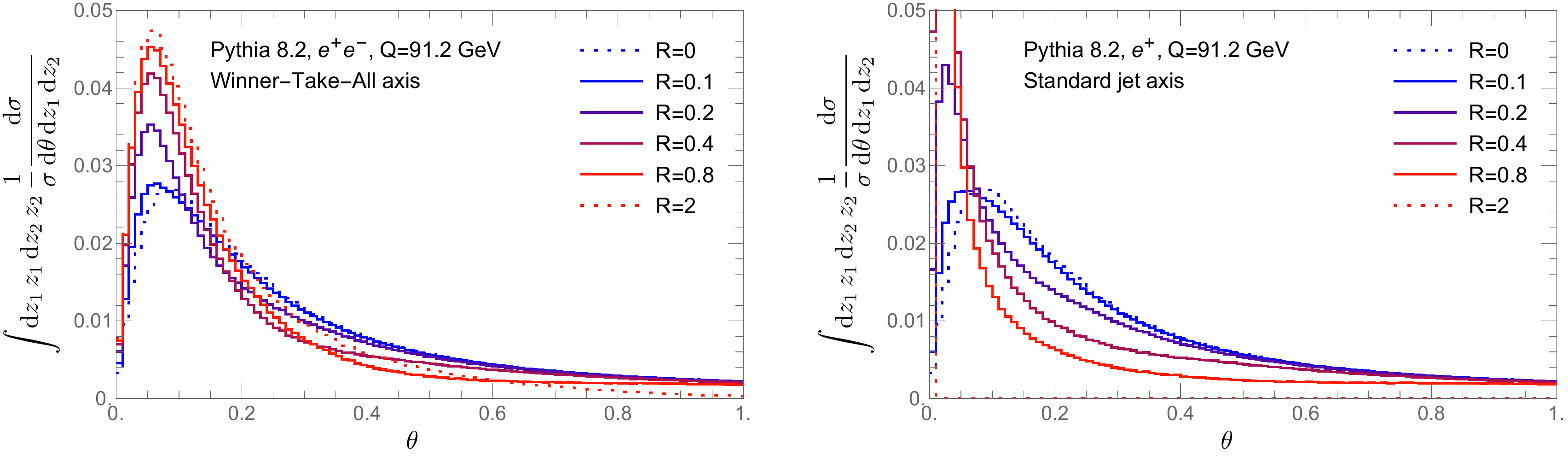} }
\caption{The dependence of the angular decorrelation $\theta$ on the jet radius $R$ for the winner-take-all axis (left) and standard jet axis (right). For small $R$ they agree, and the differences for large $R$ are consistent with our framework, as discussed in the text.}\label{fig:Pythia}
\end{figure*}

We performed independent calculations in both impact parameter and transverse momentum space, providing a cross-check.  Writing $J_{i}^{\axis}(z,\vecb b,QR,\mu,\ze) =\sum_k a_s^k J_{i}^{[k]\,\axis}(z,\vecb b ,QR,\mu,\ze)$ with $a_s=g_s^2/(4\pi)^2$, our result for the case $\theta \sim R$ (or equivalently $|\vecb b|QR \sim 1$) is given by 
\begin{widetext}
\begin{align}\label{eq:TMDjetResultb}
  J_{i}^{[0]\,\axis}(z,\vecb b,QR,\mu,\ze) &= \de(1-z), \\
  J_{i}^{[1]\,\axis}(z,\vecb b, QR,\mu,\ze) 
  &= 
\de(1-z)\bigl[2\cC'_i\,L_R
-\cC_i L_\mu^2+2\cC_i L_\mu \mathbf{l}_\ze+2\tilde{d}_i^\axis(\vecb bQR)\bigr]
+2\Bigl[ \sum_j c_{ji}\, p_{ji}(z)\Bigr]
  \nn \\ & \quad \times
  \bigl[L_R-L_\mu-2\ln (1-z)+\tfrac14 |\vecb b|^2 Q^2 R^2 (1-z)^2\,_2F_3(\{1,1\},\{2,2,2\};-\tfrac14 |\vecb b|^2 Q^2 R^2 (1-z)^2)\bigr],
\nn
\end{align}
\end{widetext}
where 
\begin{align}
  L_R = \ln\Big(\frac{\mu^2}{Q^2R^2}\Big), &\quad  \mathbf{l}_\ze=\ln\frac{\mu^2}{\ze},
  &L_\mu= \LB.
\end{align}
The group theory factors are given by
$\cC_q =c_{qq}=c_{gq}= C_F$, $ \cC_g = \tfrac12 c_{gg}=C_A$, $\cC'_q = \tfrac32 C_F$, $2\cC'_g = \bt_0 = \tfrac{11}{3} C_A- \tfrac{4}{3} n_f T_F$, $c_{qg}=2n_fT_F$.
We use the convention for the well-known splitting functions $p_{ij}$ of Refs.~\cite{Vogt:2004mw,Moch:2004pa}, including their plus prescription at $z=1$ when a splitting function multiplies $\ln(1-z)$.

We report $\tilde d^{\rm axis}$ of \eq{TMDjetResultb} in momentum space $q = |\vecb q|$, because they are much more compact, deferring complete expressions in $\vecb b$-space to Ref.~\cite{nos}. These are given by 
\begin{align}\label{eq:dFunction}
  d_q^\WTA &= \frac{C_F}{\pi}\bigg\{\de(q^2)\Big(\frac{7}{2}-\frac{5\pi^2}{12}-2\ln^2 2\Big)
  \nn \\ &\quad
  -\Big(\frac{3}{2}-2\ln 2\Big)\LzeroQTR-\LoneQTR
  \nn\\ &\quad
  +\thh\Big(\frac{QR}{2}-q\Big)\frac{1}{q^2}\Big[3\frac{q}{QR}+2\ln\Big(1-\frac{q}{QR}\Big)\Big]\bigg\}\nn,\\
  d_g^\WTA &= \frac{1}{\pi}\bigg\{\de(q^2)\bigg[C_A\Big(\frac{131}{36}-\frac{5\pi^2}{12}-2\ln^2 2\Big)-\frac{17}{18}n_fT_R
  \bigg]
  \nn\\ &\quad
   -\Big(\frac{\bt_0}{2}-2C_A\ln 2\Big)\LzeroQTR-C_A\LoneQTR
   \nn\\ &\quad
  +\thh\Big(\frac{QR}{2}-q\Big)\frac{1}{q^2}\bigg[2n_fT_R\bigg(-\frac{q}{QR}+\Big(\frac{q}{QR}\Big)^2
  \nn\\ &\quad
  -\frac{2}{3}\Big(\frac{q}{QR}\Big)^3\bigg)+C_A\bigg(4\frac{q}{QR}-\Big(\frac{q}{QR}\Big)^2
  \nn\\ &\quad
  +\frac{2}{3}\Big(\frac{q}{QR}\Big)^3+2\ln\Big(1-\frac{q}{QR}\Big)\bigg)\bigg]\bigg\}.
\end{align}
The auxiliary plus distributions that enter here are
\begin{align}
  \cL^n_+(q,q_0) = \thh(q_0-q)\frac{1}{q_0^2}\biggl[\frac{\ln^n (q^2/q_0^2)}{q^2/q_0^2}\biggr]_+.
\end{align}

Interestingly, as a direct calculation reveals, the corresponding $d_i^{\rm SJA}$ for the standard jet axis are simply the $q\gg QR$ limits of \eq{dFunction}. This is consistent with our earlier statement that for $\thh\gg R$ the TMD jet functions are independent of the jet axis. We have also verified \eq{smallTheta}, describing the factorization in this limit.

Finally, it is interesting to take the $q \ll QR$ limit of \eq{TMDjetResultb} for the WTA, which yields
\begin{align}
{\mathscr{J}}_i^{[1]\WTA}(\vecb{b},\mu,\ze)&=
  2\Bigl\{ 
N_i+ L_\mu\Bigl[{\cal C}_i^\prime +{\cal C}_i\Bigl(\mathbf{l}_\ze- \frac{1}{2}L_\mu\Bigr)\Bigr]\Bigr\}\,,
\nn \\ 
N_g&= C_A\Big(\frac{131}{36}\!-\!\frac{5\pi^2}{12}\Big)\!-\!\frac{17}{18} n_f T_R\!-\!\beta_0\ln 2
\,,\nn \\
N_q&= C_F \Big(\frac{7}{2}-\frac{5\pi^2}{12}-3\ln 2\Big)
\,.\end{align}
As anticipated in \eq{wta}, when $\thh \ll R$ the dependence on the momentum fraction $z$ is trivial and the result becomes independent of $R$.

{\it Discussion and conclusions --}
We propose to use the measurements of jet momenta to study TMDs and for this purpose we have investigated the factorization of the cross section and calculated the TMD jet functions at NLO. Although most of the discussion is centered on the momentum decorrelation in $e^+e^-$ collisions, our studies reveal how to define jet TMDs that share the same evolution as TMD fragmentation functions. A particularly promising application is represented by  SIDIS experiments, for which the factorization theorem is 
\begin{align} \label{eq:SIDIS}
  \frac{\df \sigma_{(e N\rightarrow e J X)}}{\df Q^2\, \df x\, \df z\, \df \vecb{q}} & = \sum_a \mathcal{H}_a(Q^2,\mu) \!\! \int
  \frac{\df \vecb{b}}{(2\pi)^2}\,e^{-\img \textit{\textbf{b}} \cdot \textit{\textbf{q}}}\,
   \\
  &\quad \times f_{a/N}(x,\vecb{b},\mu,\ze)\, J_q^{\rm axis}(z,\vecb{b},QR,\mu,\ze),
\nn \end{align}
enabling  a clean extraction of the nonperturbative TMD parton distributions from this process. This factorization only holds for all $R$ when the WTA axis is used.

Our predictions show how the factorization of the cross section using TMD distributions
 depends on the size of the jet radius. To illustrate this effect and provide additional evidence, we show in \fig{Pythia} the angular decorrelation obtained from Pythia 8.2~\cite{Sjostrand:2014zea} for various jet radii. 
 We have included the fragmentation limit $(R=0)$ and the case where all particles are clustered until there are two jets $(R= 2)$ for completeness. By integrating over $z_1$ and $z_2$, taking the second Mellin moment, the  dependence on fragmentation for $R=0$ is removed by the momentum sum rule.
 First of all we note that for small $R$, the distributions are indeed the same for the WTA and SJA. For the WTA the dependence on $R$ is fairly mild, whereas for the SJA the distribution blows up at small $\theta$ for large $R$.
 This is consistent with our finding that only for the WTA axis the cross section is well-behaved in the large $R$ limit. 
 We have also investigated the size of hadronization effects for the WTA axis, finding that these are smaller for larger values of $R$ (even though we are using the momentum sum rule). 
  As $\theta \ll R$ is the regime relevant for constraining non-perturbative physics in the initial state from \eq{SIDIS}, the smallness of hadronization effects in the final state is an important benefit.
   Last of all, we point out that one may consider jets defined solely on charged particles to overcome the limited angular resolution of calorimetry, by e.g.~using the computing framework developed in Ref.~\cite{Chang:2013rca}. A dedicated phenomenological study is forthcoming~\cite{nos}.

{\it Acknowledgements --}
We thank D.~Neill and A.~Vladimirov for feedback on this manuscript. D.G.R.~and I.S.~are supported by the Spanish MECD grant FPA2016-75654-C2-2-P and the group UPARCOS. D.G.R.~acknowledges the support of the Universidad Complutense de Madrid through the predoctoral grant CT17/17-CT18/17. W.W.~and L.Z.~are supported by ERC grant ERC-STG-2015-677323. W.W.~also acknowledges support by the D-ITP consortium, a program of the Netherlands Organization for Scientific Research (NWO) that is funded by the Dutch Ministry of Education, Culture and Science (OCW). This article  is based upon work from COST Action CA16201 PARTICLEFACE, supported by COST (European Cooperation in Science and Technology).

\bibliography{dijet_tmd}

\begin{thebibliography}{60}%
\makeatletter
\providecommand \@ifxundefined [1]{%
 \@ifx{#1\undefined}
}%
\providecommand \@ifnum [1]{%
 \ifnum #1\expandafter \@firstoftwo
 \else \expandafter \@secondoftwo
 \fi
}%
\providecommand \@ifx [1]{%
 \ifx #1\expandafter \@firstoftwo
 \else \expandafter \@secondoftwo
 \fi
}%
\providecommand \natexlab [1]{#1}%
\providecommand \enquote  [1]{``#1''}%
\providecommand \bibnamefont  [1]{#1}%
\providecommand \bibfnamefont [1]{#1}%
\providecommand \citenamefont [1]{#1}%
\providecommand \href@noop [0]{\@secondoftwo}%
\providecommand \href [0]{\begingroup \@sanitize@url \@href}%
\providecommand \@href[1]{\@@startlink{#1}\@@href}%
\providecommand \@@href[1]{\endgroup#1\@@endlink}%
\providecommand \@sanitize@url [0]{\catcode `\\12\catcode `\$12\catcode
  `\&12\catcode `\#12\catcode `\^12\catcode `\_12\catcode `\%12\relax}%
\providecommand \@@startlink[1]{}%
\providecommand \@@endlink[0]{}%
\providecommand \url  [0]{\begingroup\@sanitize@url \@url }%
\providecommand \@url [1]{\endgroup\@href {#1}{\urlprefix }}%
\providecommand \urlprefix  [0]{URL }%
\providecommand \Eprint [0]{\href }%
\providecommand \doibase [0]{http://dx.doi.org/}%
\providecommand \selectlanguage [0]{\@gobble}%
\providecommand \bibinfo  [0]{\@secondoftwo}%
\providecommand \bibfield  [0]{\@secondoftwo}%
\providecommand \translation [1]{[#1]}%
\providecommand \BibitemOpen [0]{}%
\providecommand \bibitemStop [0]{}%
\providecommand \bibitemNoStop [0]{.\EOS\space}%
\providecommand \EOS [0]{\spacefactor3000\relax}%
\providecommand \BibitemShut  [1]{\csname bibitem#1\endcsname}%
\let\auto@bib@innerbib\@empty
\bibitem [{\citenamefont {Collins}\ and\ \citenamefont
  {Soper}(1981)}]{Collins:1981uk}%
  \BibitemOpen
  \bibfield  {author} {\bibinfo {author} {\bibfnamefont {J.~C.}\ \bibnamefont
  {Collins}}\ and\ \bibinfo {author} {\bibfnamefont {D.~E.}\ \bibnamefont
  {Soper}},\ }\href {\doibase 10.1016/0550-3213(81)90339-4} {\bibfield
  {journal} {\bibinfo  {journal} {Nucl. Phys.}\ }\textbf {\bibinfo {volume}
  {B193}},\ \bibinfo {pages} {381} (\bibinfo {year} {1981})},\ \bibinfo {note}
  {[Erratum: Nucl. Phys.B213,545(1983)]}\BibitemShut {NoStop}%
\bibitem [{\citenamefont {Collins}\ \emph {et~al.}(1985)\citenamefont
  {Collins}, \citenamefont {Soper},\ and\ \citenamefont
  {Sterman}}]{Collins:1984kg}%
  \BibitemOpen
  \bibfield  {author} {\bibinfo {author} {\bibfnamefont {J.~C.}\ \bibnamefont
  {Collins}}, \bibinfo {author} {\bibfnamefont {D.~E.}\ \bibnamefont {Soper}},
  \ and\ \bibinfo {author} {\bibfnamefont {G.~F.}\ \bibnamefont {Sterman}},\
  }\href {\doibase 10.1016/0550-3213(85)90479-1} {\bibfield  {journal}
  {\bibinfo  {journal} {Nucl. Phys.}\ }\textbf {\bibinfo {volume} {B250}},\
  \bibinfo {pages} {199} (\bibinfo {year} {1985})}\BibitemShut {NoStop}%
\bibitem [{\citenamefont {Meng}\ \emph {et~al.}(1996)\citenamefont {Meng},
  \citenamefont {Olness},\ and\ \citenamefont {Soper}}]{Meng:1995yn}%
  \BibitemOpen
  \bibfield  {author} {\bibinfo {author} {\bibfnamefont {R.}~\bibnamefont
  {Meng}}, \bibinfo {author} {\bibfnamefont {F.~I.}\ \bibnamefont {Olness}}, \
  and\ \bibinfo {author} {\bibfnamefont {D.~E.}\ \bibnamefont {Soper}},\ }\href
  {\doibase 10.1103/PhysRevD.54.1919} {\bibfield  {journal} {\bibinfo
  {journal} {Phys. Rev.}\ }\textbf {\bibinfo {volume} {D54}},\ \bibinfo {pages}
  {1919} (\bibinfo {year} {1996})},\ \Eprint
  {http://arxiv.org/abs/hep-ph/9511311} {arXiv:hep-ph/9511311 [hep-ph]}
  \BibitemShut {NoStop}%
\bibitem [{\citenamefont {Becher}\ and\ \citenamefont
  {Neubert}(2011)}]{Becher:2010tm}%
  \BibitemOpen
  \bibfield  {author} {\bibinfo {author} {\bibfnamefont {T.}~\bibnamefont
  {Becher}}\ and\ \bibinfo {author} {\bibfnamefont {M.}~\bibnamefont
  {Neubert}},\ }\href {\doibase 10.1140/epjc/s10052-011-1665-7} {\bibfield
  {journal} {\bibinfo  {journal} {Eur. Phys. J.}\ }\textbf {\bibinfo {volume}
  {C71}},\ \bibinfo {pages} {1665} (\bibinfo {year} {2011})},\ \Eprint
  {http://arxiv.org/abs/1007.4005} {arXiv:1007.4005 [hep-ph]} \BibitemShut
  {NoStop}%
\bibitem [{\citenamefont {Collins}(2013)}]{Collins:2011zzd}%
  \BibitemOpen
  \bibfield  {author} {\bibinfo {author} {\bibfnamefont {J.}~\bibnamefont
  {Collins}},\ }\href {http://www.cambridge.org/de/knowledge/isbn/item5756723}
  {\emph {\bibinfo {title} {{Foundations of perturbative QCD}}}}\ (\bibinfo
  {publisher} {Cambridge University Press},\ \bibinfo {year}
  {2013})\BibitemShut {NoStop}%
\bibitem [{\citenamefont {Echevarria}\ \emph {et~al.}(2012)\citenamefont
  {Echevarria}, \citenamefont {Idilbi},\ and\ \citenamefont
  {Scimemi}}]{GarciaEchevarria:2011rb}%
  \BibitemOpen
  \bibfield  {author} {\bibinfo {author} {\bibfnamefont {M.~G.}\ \bibnamefont
  {Echevarria}}, \bibinfo {author} {\bibfnamefont {A.}~\bibnamefont {Idilbi}},
  \ and\ \bibinfo {author} {\bibfnamefont {I.}~\bibnamefont {Scimemi}},\ }\href
  {\doibase 10.1007/JHEP07(2012)002} {\bibfield  {journal} {\bibinfo  {journal}
  {JHEP}\ }\textbf {\bibinfo {volume} {07}},\ \bibinfo {pages} {002} (\bibinfo
  {year} {2012})},\ \Eprint {http://arxiv.org/abs/1111.4996} {arXiv:1111.4996
  [hep-ph]} \BibitemShut {NoStop}%
\bibitem [{\citenamefont {Chiu}\ \emph {et~al.}(2012)\citenamefont {Chiu},
  \citenamefont {Jain}, \citenamefont {Neill},\ and\ \citenamefont
  {Rothstein}}]{Chiu:2012ir}%
  \BibitemOpen
  \bibfield  {author} {\bibinfo {author} {\bibfnamefont {J.-Y.}\ \bibnamefont
  {Chiu}}, \bibinfo {author} {\bibfnamefont {A.}~\bibnamefont {Jain}}, \bibinfo
  {author} {\bibfnamefont {D.}~\bibnamefont {Neill}}, \ and\ \bibinfo {author}
  {\bibfnamefont {I.~Z.}\ \bibnamefont {Rothstein}},\ }\href {\doibase
  10.1007/JHEP05(2012)084} {\bibfield  {journal} {\bibinfo  {journal} {JHEP}\
  }\textbf {\bibinfo {volume} {05}},\ \bibinfo {pages} {084} (\bibinfo {year}
  {2012})},\ \Eprint {http://arxiv.org/abs/1202.0814} {arXiv:1202.0814
  [hep-ph]} \BibitemShut {NoStop}%
\bibitem [{\citenamefont {Echevarria}\ \emph {et~al.}(2014)\citenamefont
  {Echevarria}, \citenamefont {Idilbi},\ and\ \citenamefont
  {Scimemi}}]{Echevarria:2014rua}%
  \BibitemOpen
  \bibfield  {author} {\bibinfo {author} {\bibfnamefont {M.~G.}\ \bibnamefont
  {Echevarria}}, \bibinfo {author} {\bibfnamefont {A.}~\bibnamefont {Idilbi}},
  \ and\ \bibinfo {author} {\bibfnamefont {I.}~\bibnamefont {Scimemi}},\ }\href
  {\doibase 10.1103/PhysRevD.90.014003} {\bibfield  {journal} {\bibinfo
  {journal} {Phys. Rev.}\ }\textbf {\bibinfo {volume} {D90}},\ \bibinfo {pages}
  {014003} (\bibinfo {year} {2014})},\ \Eprint {http://arxiv.org/abs/1402.0869}
  {arXiv:1402.0869 [hep-ph]} \BibitemShut {NoStop}%
\bibitem [{\citenamefont {Boer}\ \emph {et~al.}(2016)\citenamefont {Boer},
  \citenamefont {Mulders}, \citenamefont {Pisano},\ and\ \citenamefont
  {Zhou}}]{Boer:2016fqd}%
  \BibitemOpen
  \bibfield  {author} {\bibinfo {author} {\bibfnamefont {D.}~\bibnamefont
  {Boer}}, \bibinfo {author} {\bibfnamefont {P.~J.}\ \bibnamefont {Mulders}},
  \bibinfo {author} {\bibfnamefont {C.}~\bibnamefont {Pisano}}, \ and\ \bibinfo
  {author} {\bibfnamefont {J.}~\bibnamefont {Zhou}},\ }\href {\doibase
  10.1007/JHEP08(2016)001} {\bibfield  {journal} {\bibinfo  {journal} {JHEP}\
  }\textbf {\bibinfo {volume} {08}},\ \bibinfo {pages} {001} (\bibinfo {year}
  {2016})},\ \Eprint {http://arxiv.org/abs/1605.07934} {arXiv:1605.07934
  [hep-ph]} \BibitemShut {NoStop}%
\bibitem [{\citenamefont {Page}(2018)}]{Page:2017pcx}%
  \BibitemOpen
  \bibfield  {author} {\bibinfo {author} {\bibfnamefont {B.}~\bibnamefont
  {Page}},\ }\href@noop {} {\bibfield  {journal} {\bibinfo  {journal} {PoS}\
  }\textbf {\bibinfo {volume} {DIS2017}},\ \bibinfo {pages} {110} (\bibinfo
  {year} {2018})}\BibitemShut {NoStop}%
\bibitem [{\citenamefont {Kang}\ \emph
  {et~al.}(2017{\natexlab{a}})\citenamefont {Kang}, \citenamefont {Liu},
  \citenamefont {Ringer},\ and\ \citenamefont {Xing}}]{Kang:2017glf}%
  \BibitemOpen
  \bibfield  {author} {\bibinfo {author} {\bibfnamefont {Z.-B.}\ \bibnamefont
  {Kang}}, \bibinfo {author} {\bibfnamefont {X.}~\bibnamefont {Liu}}, \bibinfo
  {author} {\bibfnamefont {F.}~\bibnamefont {Ringer}}, \ and\ \bibinfo {author}
  {\bibfnamefont {H.}~\bibnamefont {Xing}},\ }\href {\doibase
  10.1007/JHEP11(2017)068} {\bibfield  {journal} {\bibinfo  {journal} {JHEP}\
  }\textbf {\bibinfo {volume} {11}},\ \bibinfo {pages} {068} (\bibinfo {year}
  {2017}{\natexlab{a}})},\ \Eprint {http://arxiv.org/abs/1705.08443}
  {arXiv:1705.08443 [hep-ph]} \BibitemShut {NoStop}%
\bibitem [{\citenamefont {Kang}\ \emph
  {et~al.}(2017{\natexlab{b}})\citenamefont {Kang}, \citenamefont {Qiu},
  \citenamefont {Ringer}, \citenamefont {Xing},\ and\ \citenamefont
  {Zhang}}]{Kang:2017yde}%
  \BibitemOpen
  \bibfield  {author} {\bibinfo {author} {\bibfnamefont {Z.-B.}\ \bibnamefont
  {Kang}}, \bibinfo {author} {\bibfnamefont {J.-W.}\ \bibnamefont {Qiu}},
  \bibinfo {author} {\bibfnamefont {F.}~\bibnamefont {Ringer}}, \bibinfo
  {author} {\bibfnamefont {H.}~\bibnamefont {Xing}}, \ and\ \bibinfo {author}
  {\bibfnamefont {H.}~\bibnamefont {Zhang}},\ }\href {\doibase
  10.1103/PhysRevLett.119.032001} {\bibfield  {journal} {\bibinfo  {journal}
  {Phys. Rev. Lett.}\ }\textbf {\bibinfo {volume} {119}},\ \bibinfo {pages}
  {032001} (\bibinfo {year} {2017}{\natexlab{b}})},\ \Eprint
  {http://arxiv.org/abs/1702.03287} {arXiv:1702.03287 [hep-ph]} \BibitemShut
  {NoStop}%
\bibitem [{\citenamefont {Salam}()}]{Salam:WTAUnpublished}%
  \BibitemOpen
  \bibfield  {author} {\bibinfo {author} {\bibfnamefont {G.}~\bibnamefont
  {Salam}},\ }\href@noop {} {\bibinfo  {journal} {Unpublished}\ }\BibitemShut
  {NoStop}%
\bibitem [{\citenamefont {Bertolini}\ \emph {et~al.}(2014)\citenamefont
  {Bertolini}, \citenamefont {Chan},\ and\ \citenamefont
  {Thaler}}]{Bertolini:2013iqa}%
  \BibitemOpen
\bibfield  {journal} {  }\bibfield  {author} {\bibinfo {author} {\bibfnamefont
  {D.}~\bibnamefont {Bertolini}}, \bibinfo {author} {\bibfnamefont
  {T.}~\bibnamefont {Chan}}, \ and\ \bibinfo {author} {\bibfnamefont
  {J.}~\bibnamefont {Thaler}},\ }\href {\doibase 10.1007/JHEP04(2014)013}
  {\bibfield  {journal} {\bibinfo  {journal} {JHEP}\ }\textbf {\bibinfo
  {volume} {04}},\ \bibinfo {pages} {013} (\bibinfo {year} {2014})},\ \Eprint
  {http://arxiv.org/abs/1310.7584} {arXiv:1310.7584 [hep-ph]} \BibitemShut
  {NoStop}%
\bibitem [{\citenamefont {Neill}\ \emph {et~al.}(2017)\citenamefont {Neill},
  \citenamefont {Scimemi},\ and\ \citenamefont {Waalewijn}}]{Neill:2016vbi}%
  \BibitemOpen
  \bibfield  {author} {\bibinfo {author} {\bibfnamefont {D.}~\bibnamefont
  {Neill}}, \bibinfo {author} {\bibfnamefont {I.}~\bibnamefont {Scimemi}}, \
  and\ \bibinfo {author} {\bibfnamefont {W.~J.}\ \bibnamefont {Waalewijn}},\
  }\href {\doibase 10.1007/JHEP04(2017)020} {\bibfield  {journal} {\bibinfo
  {journal} {JHEP}\ }\textbf {\bibinfo {volume} {04}},\ \bibinfo {pages} {020}
  (\bibinfo {year} {2017})},\ \Eprint {http://arxiv.org/abs/1612.04817}
  {arXiv:1612.04817 [hep-ph]} \BibitemShut {NoStop}%
\bibitem [{\citenamefont {Abazov}\ \emph {et~al.}(2005)\citenamefont {Abazov}
  \emph {et~al.}}]{Abazov:2004hm}%
  \BibitemOpen
  \bibfield  {author} {\bibinfo {author} {\bibfnamefont {V.~M.}\ \bibnamefont
  {Abazov}} \emph {et~al.} (\bibinfo {collaboration} {D0}),\ }\href {\doibase
  10.1103/PhysRevLett.94.221801} {\bibfield  {journal} {\bibinfo  {journal}
  {Phys. Rev. Lett.}\ }\textbf {\bibinfo {volume} {94}},\ \bibinfo {pages}
  {221801} (\bibinfo {year} {2005})},\ \Eprint
  {http://arxiv.org/abs/hep-ex/0409040} {arXiv:hep-ex/0409040 [hep-ex]}
  \BibitemShut {NoStop}%
\bibitem [{\citenamefont {Adams}\ \emph {et~al.}(2006)\citenamefont {Adams}
  \emph {et~al.}}]{Adams:2006yt}%
  \BibitemOpen
  \bibfield  {author} {\bibinfo {author} {\bibfnamefont {J.}~\bibnamefont
  {Adams}} \emph {et~al.} (\bibinfo {collaboration} {STAR}),\ }\href {\doibase
  10.1103/PhysRevLett.97.162301} {\bibfield  {journal} {\bibinfo  {journal}
  {Phys. Rev. Lett.}\ }\textbf {\bibinfo {volume} {97}},\ \bibinfo {pages}
  {162301} (\bibinfo {year} {2006})},\ \Eprint
  {http://arxiv.org/abs/nucl-ex/0604018} {arXiv:nucl-ex/0604018} \BibitemShut
  {NoStop}%
\bibitem [{\citenamefont {Khachatryan}\ \emph {et~al.}(2016)\citenamefont
  {Khachatryan} \emph {et~al.}}]{Khachatryan:2016hkr}%
  \BibitemOpen
  \bibfield  {author} {\bibinfo {author} {\bibfnamefont {V.}~\bibnamefont
  {Khachatryan}} \emph {et~al.} (\bibinfo {collaboration} {CMS}),\ }\href
  {\doibase 10.1140/epjc/s10052-016-4346-8} {\bibfield  {journal} {\bibinfo
  {journal} {Eur. Phys. J.}\ }\textbf {\bibinfo {volume} {C76}},\ \bibinfo
  {pages} {536} (\bibinfo {year} {2016})},\ \Eprint
  {http://arxiv.org/abs/1602.04384} {arXiv:1602.04384 [hep-ex]} \BibitemShut
  {NoStop}%
\bibitem [{\citenamefont {Aaboud}\ \emph {et~al.}(2018)\citenamefont {Aaboud}
  \emph {et~al.}}]{Aaboud:2018hie}%
  \BibitemOpen
  \bibfield  {author} {\bibinfo {author} {\bibfnamefont {M.}~\bibnamefont
  {Aaboud}} \emph {et~al.} (\bibinfo {collaboration} {ATLAS}),\ }\href@noop {}
  {\  (\bibinfo {year} {2018})},\ \Eprint {http://arxiv.org/abs/1805.04691}
  {arXiv:1805.04691 [hep-ex]} \BibitemShut {NoStop}%
\bibitem [{\citenamefont {Banfi}\ \emph {et~al.}(2008)\citenamefont {Banfi},
  \citenamefont {Dasgupta},\ and\ \citenamefont {Delenda}}]{Banfi:2008qs}%
  \BibitemOpen
  \bibfield  {author} {\bibinfo {author} {\bibfnamefont {A.}~\bibnamefont
  {Banfi}}, \bibinfo {author} {\bibfnamefont {M.}~\bibnamefont {Dasgupta}}, \
  and\ \bibinfo {author} {\bibfnamefont {Y.}~\bibnamefont {Delenda}},\ }\href
  {\doibase 10.1016/j.physletb.2008.05.065} {\bibfield  {journal} {\bibinfo
  {journal} {Phys. Lett.}\ }\textbf {\bibinfo {volume} {B665}},\ \bibinfo
  {pages} {86} (\bibinfo {year} {2008})},\ \Eprint
  {http://arxiv.org/abs/0804.3786} {arXiv:0804.3786 [hep-ph]} \BibitemShut
  {NoStop}%
\bibitem [{\citenamefont {Sun}\ \emph {et~al.}(2014)\citenamefont {Sun},
  \citenamefont {Yuan},\ and\ \citenamefont {Yuan}}]{Sun:2014gfa}%
  \BibitemOpen
  \bibfield  {author} {\bibinfo {author} {\bibfnamefont {P.}~\bibnamefont
  {Sun}}, \bibinfo {author} {\bibfnamefont {C.~P.}\ \bibnamefont {Yuan}}, \
  and\ \bibinfo {author} {\bibfnamefont {F.}~\bibnamefont {Yuan}},\ }\href
  {\doibase 10.1103/PhysRevLett.113.232001} {\bibfield  {journal} {\bibinfo
  {journal} {Phys. Rev. Lett.}\ }\textbf {\bibinfo {volume} {113}},\ \bibinfo
  {pages} {232001} (\bibinfo {year} {2014})},\ \Eprint
  {http://arxiv.org/abs/1405.1105} {arXiv:1405.1105 [hep-ph]} \BibitemShut
  {NoStop}%
\bibitem [{\citenamefont {Chen}\ \emph {et~al.}(2018)\citenamefont {Chen},
  \citenamefont {Qin}, \citenamefont {Wei}, \citenamefont {Xiao},\ and\
  \citenamefont {Zhang}}]{Chen:2016jfu}%
  \BibitemOpen
  \bibfield  {author} {\bibinfo {author} {\bibfnamefont {L.}~\bibnamefont
  {Chen}}, \bibinfo {author} {\bibfnamefont {G.-Y.}\ \bibnamefont {Qin}},
  \bibinfo {author} {\bibfnamefont {S.-Y.}\ \bibnamefont {Wei}}, \bibinfo
  {author} {\bibfnamefont {B.-W.}\ \bibnamefont {Xiao}}, \ and\ \bibinfo
  {author} {\bibfnamefont {H.-Z.}\ \bibnamefont {Zhang}},\ }\href {\doibase
  10.1016/j.physletb.2018.06.002} {\bibfield  {journal} {\bibinfo  {journal}
  {Phys. Lett.}\ }\textbf {\bibinfo {volume} {B782}},\ \bibinfo {pages} {773}
  (\bibinfo {year} {2018})},\ \Eprint {http://arxiv.org/abs/1612.04202}
  {arXiv:1612.04202 [hep-ph]} \BibitemShut {NoStop}%
\bibitem [{\citenamefont {Cacciari}\ \emph {et~al.}(2008)\citenamefont
  {Cacciari}, \citenamefont {Salam},\ and\ \citenamefont
  {Soyez}}]{Cacciari:2008gp}%
  \BibitemOpen
  \bibfield  {author} {\bibinfo {author} {\bibfnamefont {M.}~\bibnamefont
  {Cacciari}}, \bibinfo {author} {\bibfnamefont {G.~P.}\ \bibnamefont {Salam}},
  \ and\ \bibinfo {author} {\bibfnamefont {G.}~\bibnamefont {Soyez}},\ }\href
  {\doibase 10.1088/1126-6708/2008/04/063} {\bibfield  {journal} {\bibinfo
  {journal} {JHEP}\ }\textbf {\bibinfo {volume} {04}},\ \bibinfo {pages} {063}
  (\bibinfo {year} {2008})},\ \Eprint {http://arxiv.org/abs/0802.1189}
  {arXiv:0802.1189 [hep-ph]} \BibitemShut {NoStop}%
\bibitem [{\citenamefont {Bassetto}\ \emph {et~al.}(1985)\citenamefont
  {Bassetto}, \citenamefont {Dalbosco}, \citenamefont {Lazzizzera},\ and\
  \citenamefont {Soldati}}]{Bassetto:1984dq}%
  \BibitemOpen
  \bibfield  {author} {\bibinfo {author} {\bibfnamefont {A.}~\bibnamefont
  {Bassetto}}, \bibinfo {author} {\bibfnamefont {M.}~\bibnamefont {Dalbosco}},
  \bibinfo {author} {\bibfnamefont {I.}~\bibnamefont {Lazzizzera}}, \ and\
  \bibinfo {author} {\bibfnamefont {R.}~\bibnamefont {Soldati}},\ }\href
  {\doibase 10.1103/PhysRevD.31.2012} {\bibfield  {journal} {\bibinfo
  {journal} {Phys. Rev.}\ }\textbf {\bibinfo {volume} {D31}},\ \bibinfo {pages}
  {2012} (\bibinfo {year} {1985})}\BibitemShut {NoStop}%
\bibitem [{\citenamefont {Hautmann}(2007)}]{Hautmann:2007uw}%
  \BibitemOpen
  \bibfield  {author} {\bibinfo {author} {\bibfnamefont {F.}~\bibnamefont
  {Hautmann}},\ }\href {\doibase 10.1016/j.physletb.2007.08.081} {\bibfield
  {journal} {\bibinfo  {journal} {Phys. Lett.}\ }\textbf {\bibinfo {volume}
  {B655}},\ \bibinfo {pages} {26} (\bibinfo {year} {2007})},\ \Eprint
  {http://arxiv.org/abs/hep-ph/0702196} {arXiv:hep-ph/0702196} \BibitemShut
  {NoStop}%
\bibitem [{\citenamefont {Cherednikov}\ and\ \citenamefont
  {Stefanis}(2008)}]{Cherednikov:2007tw}%
  \BibitemOpen
  \bibfield  {author} {\bibinfo {author} {\bibfnamefont {I.~O.}\ \bibnamefont
  {Cherednikov}}\ and\ \bibinfo {author} {\bibfnamefont {N.~G.}\ \bibnamefont
  {Stefanis}},\ }\href {\doibase 10.1103/PhysRevD.77.094001} {\bibfield
  {journal} {\bibinfo  {journal} {Phys. Rev.}\ }\textbf {\bibinfo {volume}
  {D77}},\ \bibinfo {pages} {094001} (\bibinfo {year} {2008})},\ \Eprint
  {http://arxiv.org/abs/0710.1955} {arXiv:0710.1955 [hep-ph]} \BibitemShut
  {NoStop}%
\bibitem [{\citenamefont {Cherednikov}\ and\ \citenamefont
  {Stefanis}(2009)}]{Cherednikov:2009wk}%
  \BibitemOpen
  \bibfield  {author} {\bibinfo {author} {\bibfnamefont {I.~O.}\ \bibnamefont
  {Cherednikov}}\ and\ \bibinfo {author} {\bibfnamefont {N.~G.}\ \bibnamefont
  {Stefanis}},\ }\href {\doibase 10.1103/PhysRevD.80.054008} {\bibfield
  {journal} {\bibinfo  {journal} {Phys. Rev.}\ }\textbf {\bibinfo {volume}
  {D80}},\ \bibinfo {pages} {054008} (\bibinfo {year} {2009})},\ \Eprint
  {http://arxiv.org/abs/0904.2727} {arXiv:0904.2727 [hep-ph]} \BibitemShut
  {NoStop}%
\bibitem [{\citenamefont {Idilbi}\ and\ \citenamefont
  {Scimemi}(2011)}]{Idilbi:2010im}%
  \BibitemOpen
  \bibfield  {author} {\bibinfo {author} {\bibfnamefont {A.}~\bibnamefont
  {Idilbi}}\ and\ \bibinfo {author} {\bibfnamefont {I.}~\bibnamefont
  {Scimemi}},\ }\href {\doibase 10.1016/j.physletb.2010.11.060} {\bibfield
  {journal} {\bibinfo  {journal} {Phys. Lett.}\ }\textbf {\bibinfo {volume}
  {B695}},\ \bibinfo {pages} {463} (\bibinfo {year} {2011})},\ \Eprint
  {http://arxiv.org/abs/1009.2776} {arXiv:1009.2776 [hep-ph]} \BibitemShut
  {NoStop}%
\bibitem [{\citenamefont {Garcia-Echevarria}\ \emph {et~al.}(2011)\citenamefont
  {Garcia-Echevarria}, \citenamefont {Idilbi},\ and\ \citenamefont
  {Scimemi}}]{GarciaEchevarria:2011md}%
  \BibitemOpen
  \bibfield  {author} {\bibinfo {author} {\bibfnamefont {M.}~\bibnamefont
  {Garcia-Echevarria}}, \bibinfo {author} {\bibfnamefont {A.}~\bibnamefont
  {Idilbi}}, \ and\ \bibinfo {author} {\bibfnamefont {I.}~\bibnamefont
  {Scimemi}},\ }\href {\doibase 10.1103/PhysRevD.84.011502} {\bibfield
  {journal} {\bibinfo  {journal} {Phys. Rev.}\ }\textbf {\bibinfo {volume}
  {D84}},\ \bibinfo {pages} {011502} (\bibinfo {year} {2011})},\ \Eprint
  {http://arxiv.org/abs/1104.0686} {arXiv:1104.0686 [hep-ph]} \BibitemShut
  {NoStop}%
\bibitem [{\citenamefont {Echevarria}\ \emph
  {et~al.}(2016{\natexlab{a}})\citenamefont {Echevarria}, \citenamefont
  {Scimemi},\ and\ \citenamefont {Vladimirov}}]{Echevarria:2015byo}%
  \BibitemOpen
  \bibfield  {author} {\bibinfo {author} {\bibfnamefont {M.~G.}\ \bibnamefont
  {Echevarria}}, \bibinfo {author} {\bibfnamefont {I.}~\bibnamefont {Scimemi}},
  \ and\ \bibinfo {author} {\bibfnamefont {A.}~\bibnamefont {Vladimirov}},\
  }\href {\doibase 10.1103/PhysRevD.93.054004} {\bibfield  {journal} {\bibinfo
  {journal} {Phys. Rev.}\ }\textbf {\bibinfo {volume} {D93}},\ \bibinfo {pages}
  {054004} (\bibinfo {year} {2016}{\natexlab{a}})},\ \Eprint
  {http://arxiv.org/abs/1511.05590} {arXiv:1511.05590 [hep-ph]} \BibitemShut
  {NoStop}%
\bibitem [{\citenamefont {{L\"ubbert}}\ \emph {et~al.}(2016)\citenamefont
  {{L\"ubbert}}, \citenamefont {Oredsson},\ and\ \citenamefont
  {Stahlhofen}}]{Luebbert:2016itl}%
  \BibitemOpen
  \bibfield  {author} {\bibinfo {author} {\bibfnamefont {T.}~\bibnamefont
  {{L\"ubbert}}}, \bibinfo {author} {\bibfnamefont {J.}~\bibnamefont
  {Oredsson}}, \ and\ \bibinfo {author} {\bibfnamefont {M.}~\bibnamefont
  {Stahlhofen}},\ }\href {\doibase 10.1007/JHEP03(2016)168} {\bibfield
  {journal} {\bibinfo  {journal} {JHEP}\ }\textbf {\bibinfo {volume} {03}},\
  \bibinfo {pages} {168} (\bibinfo {year} {2016})},\ \Eprint
  {http://arxiv.org/abs/1602.01829} {arXiv:1602.01829 [hep-ph]} \BibitemShut
  {NoStop}%
\bibitem [{\citenamefont {Li}\ \emph {et~al.}(2016)\citenamefont {Li},
  \citenamefont {Neill},\ and\ \citenamefont {Zhu}}]{Li:2016axz}%
  \BibitemOpen
  \bibfield  {author} {\bibinfo {author} {\bibfnamefont {Y.}~\bibnamefont
  {Li}}, \bibinfo {author} {\bibfnamefont {D.}~\bibnamefont {Neill}}, \ and\
  \bibinfo {author} {\bibfnamefont {H.~X.}\ \bibnamefont {Zhu}},\ }\href@noop
  {} {\bibfield  {journal} {\bibinfo  {journal} {Submitted to: Phys. Rev. D}\ }
  (\bibinfo {year} {2016})},\ \Eprint {http://arxiv.org/abs/1604.00392}
  {arXiv:1604.00392 [hep-ph]} \BibitemShut {NoStop}%
\bibitem [{\citenamefont {Li}\ and\ \citenamefont {Zhu}(2017)}]{Li:2016ctv}%
  \BibitemOpen
  \bibfield  {author} {\bibinfo {author} {\bibfnamefont {Y.}~\bibnamefont
  {Li}}\ and\ \bibinfo {author} {\bibfnamefont {H.~X.}\ \bibnamefont {Zhu}},\
  }\href {\doibase 10.1103/PhysRevLett.118.022004} {\bibfield  {journal}
  {\bibinfo  {journal} {Phys. Rev. Lett.}\ }\textbf {\bibinfo {volume} {118}},\
  \bibinfo {pages} {022004} (\bibinfo {year} {2017})},\ \Eprint
  {http://arxiv.org/abs/1604.01404} {arXiv:1604.01404 [hep-ph]} \BibitemShut
  {NoStop}%
\bibitem [{\citenamefont {Vladimirov}(2017)}]{Vladimirov:2016dll}%
  \BibitemOpen
  \bibfield  {author} {\bibinfo {author} {\bibfnamefont {A.~A.}\ \bibnamefont
  {Vladimirov}},\ }\href {\doibase 10.1103/PhysRevLett.118.062001} {\bibfield
  {journal} {\bibinfo  {journal} {Phys. Rev. Lett.}\ }\textbf {\bibinfo
  {volume} {118}},\ \bibinfo {pages} {062001} (\bibinfo {year} {2017})},\
  \Eprint {http://arxiv.org/abs/1610.05791} {arXiv:1610.05791 [hep-ph]}
  \BibitemShut {NoStop}%
\bibitem [{\citenamefont {Catani}\ \emph {et~al.}(2001)\citenamefont {Catani},
  \citenamefont {de~Florian},\ and\ \citenamefont {Grazzini}}]{Catani:2000vq}%
  \BibitemOpen
  \bibfield  {author} {\bibinfo {author} {\bibfnamefont {S.}~\bibnamefont
  {Catani}}, \bibinfo {author} {\bibfnamefont {D.}~\bibnamefont {de~Florian}},
  \ and\ \bibinfo {author} {\bibfnamefont {M.}~\bibnamefont {Grazzini}},\
  }\href {\doibase 10.1016/S0550-3213(00)00617-9} {\bibfield  {journal}
  {\bibinfo  {journal} {Nucl. Phys.}\ }\textbf {\bibinfo {volume} {B596}},\
  \bibinfo {pages} {299} (\bibinfo {year} {2001})},\ \Eprint
  {http://arxiv.org/abs/hep-ph/0008184} {arXiv:hep-ph/0008184 [hep-ph]}
  \BibitemShut {NoStop}%
\bibitem [{\citenamefont {Aybat}\ and\ \citenamefont
  {Rogers}(2011)}]{Aybat:2011zv}%
  \BibitemOpen
  \bibfield  {author} {\bibinfo {author} {\bibfnamefont {S.~M.}\ \bibnamefont
  {Aybat}}\ and\ \bibinfo {author} {\bibfnamefont {T.~C.}\ \bibnamefont
  {Rogers}},\ }\href {\doibase 10.1103/PhysRevD.83.114042} {\bibfield
  {journal} {\bibinfo  {journal} {Phys. Rev.}\ }\textbf {\bibinfo {volume}
  {D83}},\ \bibinfo {pages} {114042} (\bibinfo {year} {2011})},\ \Eprint
  {http://arxiv.org/abs/1101.5057} {arXiv:1101.5057 [hep-ph]} \BibitemShut
  {NoStop}%
\bibitem [{\citenamefont {Echevarria}\ \emph {et~al.}(2013)\citenamefont
  {Echevarria}, \citenamefont {Idilbi}, \citenamefont {Schafer},\ and\
  \citenamefont {Scimemi}}]{pippo}%
  \BibitemOpen
  \bibfield  {author} {\bibinfo {author} {\bibfnamefont {M.~G.}\ \bibnamefont
  {Echevarria}}, \bibinfo {author} {\bibfnamefont {A.}~\bibnamefont {Idilbi}},
  \bibinfo {author} {\bibfnamefont {A.}~\bibnamefont {Schafer}}, \ and\
  \bibinfo {author} {\bibfnamefont {I.}~\bibnamefont {Scimemi}},\ }\href
  {\doibase 10.1140/epjc/s10052-013-2636-y} {\bibfield  {journal} {\bibinfo
  {journal} {Eur. Phys. J.}\ }\textbf {\bibinfo {volume} {C73}},\ \bibinfo
  {pages} {2636} (\bibinfo {year} {2013})},\ \Eprint
  {http://arxiv.org/abs/1208.1281} {arXiv:1208.1281 [hep-ph]} \BibitemShut
  {NoStop}%
\bibitem [{\citenamefont {Catani}\ \emph {et~al.}(2014)\citenamefont {Catani},
  \citenamefont {Cieri}, \citenamefont {de~Florian}, \citenamefont {Ferrera},\
  and\ \citenamefont {Grazzini}}]{Catani:2013tia}%
  \BibitemOpen
  \bibfield  {author} {\bibinfo {author} {\bibfnamefont {S.}~\bibnamefont
  {Catani}}, \bibinfo {author} {\bibfnamefont {L.}~\bibnamefont {Cieri}},
  \bibinfo {author} {\bibfnamefont {D.}~\bibnamefont {de~Florian}}, \bibinfo
  {author} {\bibfnamefont {G.}~\bibnamefont {Ferrera}}, \ and\ \bibinfo
  {author} {\bibfnamefont {M.}~\bibnamefont {Grazzini}},\ }\href {\doibase
  10.1016/j.nuclphysb.2014.02.011} {\bibfield  {journal} {\bibinfo  {journal}
  {Nucl. Phys.}\ }\textbf {\bibinfo {volume} {B881}},\ \bibinfo {pages} {414}
  (\bibinfo {year} {2014})},\ \Eprint {http://arxiv.org/abs/1311.1654}
  {arXiv:1311.1654 [hep-ph]} \BibitemShut {NoStop}%
\bibitem [{\citenamefont {Scimemi}\ and\ \citenamefont
  {Vladimirov}(2018)}]{Scimemi:2018xaf}%
  \BibitemOpen
  \bibfield  {author} {\bibinfo {author} {\bibfnamefont {I.}~\bibnamefont
  {Scimemi}}\ and\ \bibinfo {author} {\bibfnamefont {A.}~\bibnamefont
  {Vladimirov}},\ }\href@noop {} {\  (\bibinfo {year} {2018})},\ \Eprint
  {http://arxiv.org/abs/1803.11089} {arXiv:1803.11089 [hep-ph]} \BibitemShut
  {NoStop}%
\bibitem [{\citenamefont {Bauer}\ \emph {et~al.}(2000)\citenamefont {Bauer},
  \citenamefont {Fleming},\ and\ \citenamefont {Luke}}]{Bauer:2000ew}%
  \BibitemOpen
  \bibfield  {author} {\bibinfo {author} {\bibfnamefont {C.~W.}\ \bibnamefont
  {Bauer}}, \bibinfo {author} {\bibfnamefont {S.}~\bibnamefont {Fleming}}, \
  and\ \bibinfo {author} {\bibfnamefont {M.~E.}\ \bibnamefont {Luke}},\ }\href
  {\doibase 10.1103/PhysRevD.63.014006} {\bibfield  {journal} {\bibinfo
  {journal} {Phys. Rev.}\ }\textbf {\bibinfo {volume} {D63}},\ \bibinfo {pages}
  {014006} (\bibinfo {year} {2000})},\ \Eprint
  {http://arxiv.org/abs/hep-ph/0005275} {arXiv:hep-ph/0005275 [hep-ph]}
  \BibitemShut {NoStop}%
\bibitem [{\citenamefont {Bauer}\ \emph {et~al.}(2001)\citenamefont {Bauer},
  \citenamefont {Fleming}, \citenamefont {Pirjol},\ and\ \citenamefont
  {Stewart}}]{Bauer:2000yr}%
  \BibitemOpen
  \bibfield  {author} {\bibinfo {author} {\bibfnamefont {C.~W.}\ \bibnamefont
  {Bauer}}, \bibinfo {author} {\bibfnamefont {S.}~\bibnamefont {Fleming}},
  \bibinfo {author} {\bibfnamefont {D.}~\bibnamefont {Pirjol}}, \ and\ \bibinfo
  {author} {\bibfnamefont {I.~W.}\ \bibnamefont {Stewart}},\ }\href {\doibase
  10.1103/PhysRevD.63.114020} {\bibfield  {journal} {\bibinfo  {journal} {Phys.
  Rev.}\ }\textbf {\bibinfo {volume} {D63}},\ \bibinfo {pages} {114020}
  (\bibinfo {year} {2001})},\ \Eprint {http://arxiv.org/abs/hep-ph/0011336}
  {arXiv:hep-ph/0011336 [hep-ph]} \BibitemShut {NoStop}%
\bibitem [{\citenamefont {Bauer}\ and\ \citenamefont
  {Stewart}(2001)}]{Bauer:2001ct}%
  \BibitemOpen
  \bibfield  {author} {\bibinfo {author} {\bibfnamefont {C.~W.}\ \bibnamefont
  {Bauer}}\ and\ \bibinfo {author} {\bibfnamefont {I.~W.}\ \bibnamefont
  {Stewart}},\ }\href {\doibase 10.1016/S0370-2693(01)00902-9} {\bibfield
  {journal} {\bibinfo  {journal} {Phys. Lett.}\ }\textbf {\bibinfo {volume}
  {B516}},\ \bibinfo {pages} {134} (\bibinfo {year} {2001})},\ \Eprint
  {http://arxiv.org/abs/hep-ph/0107001} {arXiv:hep-ph/0107001 [hep-ph]}
  \BibitemShut {NoStop}%
\bibitem [{\citenamefont {Bauer}\ \emph {et~al.}(2002)\citenamefont {Bauer},
  \citenamefont {Pirjol},\ and\ \citenamefont {Stewart}}]{Bauer:2001yt}%
  \BibitemOpen
  \bibfield  {author} {\bibinfo {author} {\bibfnamefont {C.~W.}\ \bibnamefont
  {Bauer}}, \bibinfo {author} {\bibfnamefont {D.}~\bibnamefont {Pirjol}}, \
  and\ \bibinfo {author} {\bibfnamefont {I.~W.}\ \bibnamefont {Stewart}},\
  }\href {\doibase 10.1103/PhysRevD.65.054022} {\bibfield  {journal} {\bibinfo
  {journal} {Phys. Rev.}\ }\textbf {\bibinfo {volume} {D65}},\ \bibinfo {pages}
  {054022} (\bibinfo {year} {2002})},\ \Eprint
  {http://arxiv.org/abs/hep-ph/0109045} {arXiv:hep-ph/0109045 [hep-ph]}
  \BibitemShut {NoStop}%
\bibitem [{\citenamefont {Kang}\ \emph {et~al.}(2016)\citenamefont {Kang},
  \citenamefont {Ringer},\ and\ \citenamefont {Vitev}}]{Kang:2016mcy}%
  \BibitemOpen
  \bibfield  {author} {\bibinfo {author} {\bibfnamefont {Z.-B.}\ \bibnamefont
  {Kang}}, \bibinfo {author} {\bibfnamefont {F.}~\bibnamefont {Ringer}}, \ and\
  \bibinfo {author} {\bibfnamefont {I.}~\bibnamefont {Vitev}},\ }\href
  {\doibase 10.1007/JHEP10(2016)125} {\bibfield  {journal} {\bibinfo  {journal}
  {JHEP}\ }\textbf {\bibinfo {volume} {10}},\ \bibinfo {pages} {125} (\bibinfo
  {year} {2016})},\ \Eprint {http://arxiv.org/abs/1606.06732} {arXiv:1606.06732
  [hep-ph]} \BibitemShut {NoStop}%
\bibitem [{\citenamefont {Dai}\ \emph {et~al.}(2016)\citenamefont {Dai},
  \citenamefont {Kim},\ and\ \citenamefont {Leibovich}}]{Dai:2016hzf}%
  \BibitemOpen
  \bibfield  {author} {\bibinfo {author} {\bibfnamefont {L.}~\bibnamefont
  {Dai}}, \bibinfo {author} {\bibfnamefont {C.}~\bibnamefont {Kim}}, \ and\
  \bibinfo {author} {\bibfnamefont {A.~K.}\ \bibnamefont {Leibovich}},\ }\href
  {\doibase 10.1103/PhysRevD.94.114023} {\bibfield  {journal} {\bibinfo
  {journal} {Phys. Rev.}\ }\textbf {\bibinfo {volume} {D94}},\ \bibinfo {pages}
  {114023} (\bibinfo {year} {2016})},\ \Eprint
  {http://arxiv.org/abs/1606.07411} {arXiv:1606.07411 [hep-ph]} \BibitemShut
  {NoStop}%
\bibitem [{\citenamefont {Echevarria}\ \emph
  {et~al.}(2016{\natexlab{b}})\citenamefont {Echevarria}, \citenamefont
  {Scimemi},\ and\ \citenamefont {Vladimirov}}]{Echevarria:2015usa}%
  \BibitemOpen
  \bibfield  {author} {\bibinfo {author} {\bibfnamefont {M.~G.}\ \bibnamefont
  {Echevarria}}, \bibinfo {author} {\bibfnamefont {I.}~\bibnamefont {Scimemi}},
  \ and\ \bibinfo {author} {\bibfnamefont {A.}~\bibnamefont {Vladimirov}},\
  }\href {\doibase 10.1103/PhysRevD.93.011502, 10.1103/PhysRevD.94.099904}
  {\bibfield  {journal} {\bibinfo  {journal} {Phys. Rev.}\ }\textbf {\bibinfo
  {volume} {D93}},\ \bibinfo {pages} {011502} (\bibinfo {year}
  {2016}{\natexlab{b}})},\ \bibinfo {note} {[Erratum: Phys.
  Rev.D94,no.9,099904(2016)]},\ \Eprint {http://arxiv.org/abs/1509.06392}
  {arXiv:1509.06392 [hep-ph]} \BibitemShut {NoStop}%
\bibitem [{\citenamefont {Echevarria}\ \emph
  {et~al.}(2016{\natexlab{c}})\citenamefont {Echevarria}, \citenamefont
  {Scimemi},\ and\ \citenamefont {Vladimirov}}]{Echevarria:2016scs}%
  \BibitemOpen
  \bibfield  {author} {\bibinfo {author} {\bibfnamefont {M.~G.}\ \bibnamefont
  {Echevarria}}, \bibinfo {author} {\bibfnamefont {I.}~\bibnamefont {Scimemi}},
  \ and\ \bibinfo {author} {\bibfnamefont {A.}~\bibnamefont {Vladimirov}},\
  }\href {\doibase 10.1007/JHEP09(2016)004} {\bibfield  {journal} {\bibinfo
  {journal} {JHEP}\ }\textbf {\bibinfo {volume} {09}},\ \bibinfo {pages} {004}
  (\bibinfo {year} {2016}{\natexlab{c}})},\ \Eprint
  {http://arxiv.org/abs/1604.07869} {arXiv:1604.07869 [hep-ph]} \BibitemShut
  {NoStop}%
\bibitem [{\citenamefont {Kang}\ \emph
  {et~al.}(2017{\natexlab{c}})\citenamefont {Kang}, \citenamefont {Ringer},\
  and\ \citenamefont {Waalewijn}}]{Kang:2017mda}%
  \BibitemOpen
  \bibfield  {author} {\bibinfo {author} {\bibfnamefont {Z.-B.}\ \bibnamefont
  {Kang}}, \bibinfo {author} {\bibfnamefont {F.}~\bibnamefont {Ringer}}, \ and\
  \bibinfo {author} {\bibfnamefont {W.~J.}\ \bibnamefont {Waalewijn}},\ }\href
  {\doibase 10.1007/JHEP07(2017)064} {\bibfield  {journal} {\bibinfo  {journal}
  {JHEP}\ }\textbf {\bibinfo {volume} {07}},\ \bibinfo {pages} {064} (\bibinfo
  {year} {2017}{\natexlab{c}})},\ \Eprint {http://arxiv.org/abs/1705.05375}
  {arXiv:1705.05375 [hep-ph]} \BibitemShut {NoStop}%
\bibitem [{\citenamefont {Becher}\ \emph {et~al.}(2016)\citenamefont {Becher},
  \citenamefont {Neubert}, \citenamefont {Rothen},\ and\ \citenamefont
  {Shao}}]{Becher:2015hka}%
  \BibitemOpen
  \bibfield  {author} {\bibinfo {author} {\bibfnamefont {T.}~\bibnamefont
  {Becher}}, \bibinfo {author} {\bibfnamefont {M.}~\bibnamefont {Neubert}},
  \bibinfo {author} {\bibfnamefont {L.}~\bibnamefont {Rothen}}, \ and\ \bibinfo
  {author} {\bibfnamefont {D.~Y.}\ \bibnamefont {Shao}},\ }\href {\doibase
  10.1103/PhysRevLett.116.192001} {\bibfield  {journal} {\bibinfo  {journal}
  {Phys. Rev. Lett.}\ }\textbf {\bibinfo {volume} {116}},\ \bibinfo {pages}
  {192001} (\bibinfo {year} {2016})},\ \Eprint
  {http://arxiv.org/abs/1508.06645} {arXiv:1508.06645 [hep-ph]} \BibitemShut
  {NoStop}%
\bibitem [{\citenamefont {Larkoski}\ \emph {et~al.}(2015)\citenamefont
  {Larkoski}, \citenamefont {Moult},\ and\ \citenamefont
  {Neill}}]{Larkoski:2015zka}%
  \BibitemOpen
  \bibfield  {author} {\bibinfo {author} {\bibfnamefont {A.~J.}\ \bibnamefont
  {Larkoski}}, \bibinfo {author} {\bibfnamefont {I.}~\bibnamefont {Moult}}, \
  and\ \bibinfo {author} {\bibfnamefont {D.}~\bibnamefont {Neill}},\ }\href
  {\doibase 10.1007/JHEP09(2015)143} {\bibfield  {journal} {\bibinfo  {journal}
  {JHEP}\ }\textbf {\bibinfo {volume} {09}},\ \bibinfo {pages} {143} (\bibinfo
  {year} {2015})},\ \Eprint {http://arxiv.org/abs/1501.04596} {arXiv:1501.04596
  [hep-ph]} \BibitemShut {NoStop}%
\bibitem [{\citenamefont {Caron-Huot}(2018)}]{Caron-Huot:2015bja}%
  \BibitemOpen
  \bibfield  {author} {\bibinfo {author} {\bibfnamefont {S.}~\bibnamefont
  {Caron-Huot}},\ }\href {\doibase 10.1007/JHEP03(2018)036} {\bibfield
  {journal} {\bibinfo  {journal} {JHEP}\ }\textbf {\bibinfo {volume} {03}},\
  \bibinfo {pages} {036} (\bibinfo {year} {2018})},\ \Eprint
  {http://arxiv.org/abs/1501.03754} {arXiv:1501.03754 [hep-ph]} \BibitemShut
  {NoStop}%
\bibitem [{\citenamefont {Dasgupta}\ and\ \citenamefont
  {Salam}(2001)}]{Dasgupta:2001sh}%
  \BibitemOpen
  \bibfield  {author} {\bibinfo {author} {\bibfnamefont {M.}~\bibnamefont
  {Dasgupta}}\ and\ \bibinfo {author} {\bibfnamefont {G.~P.}\ \bibnamefont
  {Salam}},\ }\href {\doibase 10.1016/S0370-2693(01)00725-0} {\bibfield
  {journal} {\bibinfo  {journal} {Phys. Lett.}\ }\textbf {\bibinfo {volume}
  {B512}},\ \bibinfo {pages} {323} (\bibinfo {year} {2001})},\ \Eprint
  {http://arxiv.org/abs/hep-ph/0104277} {arXiv:hep-ph/0104277} \BibitemShut
  {NoStop}%
\bibitem [{\citenamefont {Moch}\ \emph {et~al.}(2005)\citenamefont {Moch},
  \citenamefont {Vermaseren},\ and\ \citenamefont {Vogt}}]{Moch:2005id}%
  \BibitemOpen
  \bibfield  {author} {\bibinfo {author} {\bibfnamefont {S.}~\bibnamefont
  {Moch}}, \bibinfo {author} {\bibfnamefont {J.~A.~M.}\ \bibnamefont
  {Vermaseren}}, \ and\ \bibinfo {author} {\bibfnamefont {A.}~\bibnamefont
  {Vogt}},\ }\href {\doibase 10.1088/1126-6708/2005/08/049} {\bibfield
  {journal} {\bibinfo  {journal} {JHEP}\ }\textbf {\bibinfo {volume} {08}},\
  \bibinfo {pages} {049} (\bibinfo {year} {2005})},\ \Eprint
  {http://arxiv.org/abs/hep-ph/0507039} {arXiv:hep-ph/0507039} \BibitemShut
  {NoStop}%
\bibitem [{\citenamefont {Gutierrez-Reyes}\ \emph {et~al.}()\citenamefont
  {Gutierrez-Reyes}, \citenamefont {Scimemi}, \citenamefont {Waalewijn},\ and\
  \citenamefont {Zoppi}}]{nos}%
  \BibitemOpen
  \bibfield  {author} {\bibinfo {author} {\bibfnamefont {D.}~\bibnamefont
  {Gutierrez-Reyes}}, \bibinfo {author} {\bibfnamefont {I.}~\bibnamefont
  {Scimemi}}, \bibinfo {author} {\bibfnamefont {W.~J.}\ \bibnamefont
  {Waalewijn}}, \ and\ \bibinfo {author} {\bibfnamefont {L.}~\bibnamefont
  {Zoppi}},\ }\href@noop {} {\bibinfo  {journal} {In preparation}\
  }\BibitemShut {NoStop}%
\bibitem [{\citenamefont {Chiu}\ \emph {et~al.}(2009)\citenamefont {Chiu},
  \citenamefont {Fuhrer}, \citenamefont {Hoang}, \citenamefont {Kelley},\ and\
  \citenamefont {Manohar}}]{Chiu:2009yx}%
  \BibitemOpen
\bibfield  {journal} {  }\bibfield  {author} {\bibinfo {author} {\bibfnamefont
  {J.-y.}\ \bibnamefont {Chiu}}, \bibinfo {author} {\bibfnamefont
  {A.}~\bibnamefont {Fuhrer}}, \bibinfo {author} {\bibfnamefont {A.~H.}\
  \bibnamefont {Hoang}}, \bibinfo {author} {\bibfnamefont {R.}~\bibnamefont
  {Kelley}}, \ and\ \bibinfo {author} {\bibfnamefont {A.~V.}\ \bibnamefont
  {Manohar}},\ }\href {\doibase 10.1103/PhysRevD.79.053007} {\bibfield
  {journal} {\bibinfo  {journal} {Phys. Rev.}\ }\textbf {\bibinfo {volume}
  {D79}},\ \bibinfo {pages} {053007} (\bibinfo {year} {2009})},\ \Eprint
  {http://arxiv.org/abs/0901.1332} {arXiv:0901.1332 [hep-ph]} \BibitemShut
  {NoStop}%
\bibitem [{\citenamefont {Manohar}\ and\ \citenamefont
  {Stewart}(2007)}]{Manohar:2006nz}%
  \BibitemOpen
  \bibfield  {author} {\bibinfo {author} {\bibfnamefont {A.~V.}\ \bibnamefont
  {Manohar}}\ and\ \bibinfo {author} {\bibfnamefont {I.~W.}\ \bibnamefont
  {Stewart}},\ }\href {\doibase 10.1103/PhysRevD.76.074002} {\bibfield
  {journal} {\bibinfo  {journal} {Phys. Rev.}\ }\textbf {\bibinfo {volume}
  {D76}},\ \bibinfo {pages} {074002} (\bibinfo {year} {2007})},\ \Eprint
  {http://arxiv.org/abs/hep-ph/0605001} {arXiv:hep-ph/0605001} \BibitemShut
  {NoStop}%
\bibitem [{\citenamefont {Vogt}\ \emph {et~al.}(2004)\citenamefont {Vogt},
  \citenamefont {Moch},\ and\ \citenamefont {Vermaseren}}]{Vogt:2004mw}%
  \BibitemOpen
  \bibfield  {author} {\bibinfo {author} {\bibfnamefont {A.}~\bibnamefont
  {Vogt}}, \bibinfo {author} {\bibfnamefont {S.}~\bibnamefont {Moch}}, \ and\
  \bibinfo {author} {\bibfnamefont {J.~A.~M.}\ \bibnamefont {Vermaseren}},\
  }\href {\doibase 10.1016/j.nuclphysb.2004.04.024} {\bibfield  {journal}
  {\bibinfo  {journal} {Nucl. Phys.}\ }\textbf {\bibinfo {volume} {B691}},\
  \bibinfo {pages} {129} (\bibinfo {year} {2004})},\ \Eprint
  {http://arxiv.org/abs/hep-ph/0404111} {arXiv:hep-ph/0404111} \BibitemShut
  {NoStop}%
\bibitem [{\citenamefont {Moch}\ \emph {et~al.}(2004)\citenamefont {Moch},
  \citenamefont {Vermaseren},\ and\ \citenamefont {Vogt}}]{Moch:2004pa}%
  \BibitemOpen
  \bibfield  {author} {\bibinfo {author} {\bibfnamefont {S.}~\bibnamefont
  {Moch}}, \bibinfo {author} {\bibfnamefont {J.~A.~M.}\ \bibnamefont
  {Vermaseren}}, \ and\ \bibinfo {author} {\bibfnamefont {A.}~\bibnamefont
  {Vogt}},\ }\href {\doibase 10.1016/j.nuclphysb.2004.03.030} {\bibfield
  {journal} {\bibinfo  {journal} {Nucl. Phys.}\ }\textbf {\bibinfo {volume}
  {B688}},\ \bibinfo {pages} {101} (\bibinfo {year} {2004})},\ \Eprint
  {http://arxiv.org/abs/hep-ph/0403192} {arXiv:hep-ph/0403192} \BibitemShut
  {NoStop}%
\bibitem [{\citenamefont {Sj{\"o}strand}\ \emph {et~al.}(2015)\citenamefont
  {Sj{\"o}strand}, \citenamefont {Ask}, \citenamefont {Christiansen},
  \citenamefont {Corke}, \citenamefont {Desai}, \citenamefont {Ilten},
  \citenamefont {Mrenna}, \citenamefont {Prestel}, \citenamefont {Rasmussen},\
  and\ \citenamefont {Skands}}]{Sjostrand:2014zea}%
  \BibitemOpen
  \bibfield  {author} {\bibinfo {author} {\bibfnamefont {T.}~\bibnamefont
  {Sj{\"o}strand}}, \bibinfo {author} {\bibfnamefont {S.}~\bibnamefont {Ask}},
  \bibinfo {author} {\bibfnamefont {J.~R.}\ \bibnamefont {Christiansen}},
  \bibinfo {author} {\bibfnamefont {R.}~\bibnamefont {Corke}}, \bibinfo
  {author} {\bibfnamefont {N.}~\bibnamefont {Desai}}, \bibinfo {author}
  {\bibfnamefont {P.}~\bibnamefont {Ilten}}, \bibinfo {author} {\bibfnamefont
  {S.}~\bibnamefont {Mrenna}}, \bibinfo {author} {\bibfnamefont
  {S.}~\bibnamefont {Prestel}}, \bibinfo {author} {\bibfnamefont {C.~O.}\
  \bibnamefont {Rasmussen}}, \ and\ \bibinfo {author} {\bibfnamefont {P.~Z.}\
  \bibnamefont {Skands}},\ }\href {\doibase 10.1016/j.cpc.2015.01.024}
  {\bibfield  {journal} {\bibinfo  {journal} {Comput. Phys. Commun.}\ }\textbf
  {\bibinfo {volume} {191}},\ \bibinfo {pages} {159} (\bibinfo {year}
  {2015})},\ \Eprint {http://arxiv.org/abs/1410.3012} {arXiv:1410.3012
  [hep-ph]} \BibitemShut {NoStop}%
\bibitem [{\citenamefont {Chang}\ \emph {et~al.}(2013)\citenamefont {Chang},
  \citenamefont {Procura}, \citenamefont {Thaler},\ and\ \citenamefont
  {Waalewijn}}]{Chang:2013rca}%
  \BibitemOpen
  \bibfield  {author} {\bibinfo {author} {\bibfnamefont {H.-M.}\ \bibnamefont
  {Chang}}, \bibinfo {author} {\bibfnamefont {M.}~\bibnamefont {Procura}},
  \bibinfo {author} {\bibfnamefont {J.}~\bibnamefont {Thaler}}, \ and\ \bibinfo
  {author} {\bibfnamefont {W.~J.}\ \bibnamefont {Waalewijn}},\ }\href {\doibase
  10.1103/PhysRevLett.111.102002} {\bibfield  {journal} {\bibinfo  {journal}
  {Phys. Rev. Lett.}\ }\textbf {\bibinfo {volume} {111}},\ \bibinfo {pages}
  {102002} (\bibinfo {year} {2013})},\ \Eprint {http://arxiv.org/abs/1303.6637}
  {arXiv:1303.6637 [hep-ph]} \BibitemShut {NoStop}%
\end{thebibliography}%

\end{document}